\documentstyle[stwol,epsfig]{article}

\def\be{\begin{equation}}
\def\ee{\end{equation}}
\def\bea{\begin{eqnarray}}
\def\eea{\end{eqnarray}}

\begin{document}

\title{\vskip-2.cm{\baselineskip14pt
\centerline{\normalsize\hfill IFT-12/96}
\centerline{\normalsize\hfill August 1996}
\centerline{\normalsize\hfill hep-ph/9611257}}
\vskip.5cm
ANALYSIS OF THE RENORMALIZATION SCHEME AMBIGUITIES IN THE QCD
CORRECTIONS TO HADRONIC DECAYS OF THE  TAU LEPTON$^{\mbox{\dag}}$
\vskip22cm{\baselineskip14pt\rm
\begin{flushleft}
$^{\mbox{\dag}}$ To appear in the {\it Proceedings of the 28th
International Conference on High Energy Physics}, Warsaw, Poland,
25--31 July 1996.
\end{flushleft}}
\vskip-23cm}

\author{P. A. R\c{A}CZKA}

\address{Institute of Theoretical Physics, Warsaw University, ul.\
Ho\.{z}a 69, PL-00-681 Warsaw, Poland}

\twocolumn[\maketitle\abstracts{
The QCD corrections to the $R_{\tau}^{12}$ moment of the invariant
mass distribution in hadronic decays of the $\tau$ are discussed. The
next-to-next-to-leading order prediction is shown to be stable with
respect to change of the renormalization scheme, provided that the
contour integral expression is used. The optimized predictions are
obtained using the principle of minimal sensitivity to select the
preferred renormalization scheme.  The optimized predictions for
$R_{\tau}^{12}$ and $R_{\tau}$ are used in a simplified fit to the
experimental data to determine the strong coupling constant and the
parameter characterizing the nonperturbative contribution.}]

Phenomenological applications of the finite order perturbative QCD
predictions involve an ambiguity resulting from the freedom of choice
of the renormalization scheme (RS).  Formally the RS dependence is of
higher order than the considered perturbative expression, but
numerically it may be quite significant.  In order to obtain reliable
predictions one should therefore carefully investigate the
possibilities to optimize the choice of the RS. Also, one should study
the stability of the optimized predictions by varying the RS
parameters in some {\it a~priori} acceptable range. (A condition on
the acceptable schemes, which is applicable to NNLO approximants, has
been recently proposed by the present author~\cite{par1}.)  Such an
analysis is very important in the case of the QCD corrections to the
hadronic $\tau$ decays, for which the characteristic energy scale is
$m_{\tau}=1.777~\mbox{GeV}$. In particular, it is interesting to what
extent the RS dependence affects the recent results of two
collaborations,~\cite{aleph93,aleph95,cleo} which used the hadronic
$\tau$ decay data to obtain a surprisingly precise value of
$\alpha_{s}$.

The strong interaction effects in $\tau$ may now be studied using the 
QCD corrections to the $R_{\tau}$ ratio~\cite{branapi}
\be
R_{\tau}=\frac{\Gamma(\tau \rightarrow \nu_{\tau} + hadrons)}
{\Gamma(\tau \rightarrow \nu_{\tau} e^{-} \overline{\nu}_{e})},
\ee
and the corrections to the $R_{\tau}^{kl}$ moments,
 defined by the relation~\cite{ledepi-kl}
\be
R^{kl}_{\tau}=\frac{1}{\Gamma_{e}} \int_{0}^{m_{\tau}^{2}}
ds\,\left(1-\frac{s}{m_{\tau}^{2}}\right)^{k}
\left(\frac{s}{m_{\tau}^{2}}\right)^{l}
\frac{d\Gamma_{ud}}{ds},
\ee
where $\Gamma_{e}$ is the electronic width of $\tau$ and  
$d\Gamma_{ud}/ds$ is the invariant mass distributions of the 
Cabbibo allowed hadronic decays of $\tau$.
For  $R_{\tau}^{kl}$  we have:
\be
R_{\tau}^{kl}=3\,|V_{ud}|^{2}\,S_{EW}\,
R^{kl}_{0}(1+\delta^{kl}_{pt}+\delta^{kl}_{m}+\delta^{kl}_{SVZ}),
\label{rtkl}
\ee
where the factor $S_{EW}=1.0194$ represents corrections from
electroweak interactions and $R_{0}^{kl}$ denotes the parton
model predictions. For $R_{\tau}$ we
have:
\be
R_{\tau}=3S_{CKM}S_{EW}(1+
\delta^{tot}_{pt}+\delta^{tot}_{m}+\delta^{tot}_{SVZ}),
\ee
where $S_{CKM}=(|V_{ud}|^{2}+|V_{us}|^{2})\approx1$.
The $\delta_{pt}$ contribution denotes the purely perturbative
QCD correction, evaluated for three massless quarks. (We have 
$\delta^{tot}_{pt}=\delta^{00}_{pt}$.) The $\delta_{m}$ contribution denotes
the correction from quark masses, which is practically negligible in the
case of $R_{\tau}^{kl}$. For $R_{\tau}$ we have~\cite{branapi}
$\delta^{tot}_{m}\approx0.009$. 
The $\delta_{SVZ}$ contribution is a nonperturbative
QCD correction calculated using the $SVZ$ approach~\cite{svz}
\be
\delta_{SVZ}=\sum_{D=4,6...} c_{D} 
\frac{O_{D}}{m^{D}_{\tau}}.
\label{svz}
\ee
The parameters $O_{D}$ in Eq.(\ref{svz}) denote vacuum expectation
 values of the gauge invariant operators of dimension $D$.  The
 $c_{D}$ coefficients are in principle power series in the strong
 coupling constant, and depend on the considered moment of the
 invariant mass distribution.

The interest in the $R^{kl}_{\tau}$ moments comes from the effort to
improve the accuracy of the determination of $\alpha_{s}$ from the
$\tau$ decay. Not unexpectedly, a major factor limiting the precision
of the QCD prediction for $R_{\tau}$ is 
 the uncertainty in the nonperturbative
contribution.   The contribution from the $D=4$ term in
the SVZ expansion for $R_{\tau}$ may be reliably expected to be
small~\cite{branapi}, since $O_{4}$ is well constrained by the sum
rules phenomenology, and the relevant coefficient function starts at
$O(\alpha_{s}^{2})$.  However, the $D=6$ contribution to $R_{\tau}$ is
not suppressed, and there is little information on the value of
$O_{6}$. It was therefore proposed~\cite{ledepi-kl} to treat $O_{D}$ as free
parameters which are to be extracted together with $\alpha_{s}$ from a
fit to the experimental data for $R_{\tau}$ and the higher moments of the
invariant mass distribution.  Of particular interest here is the
$R^{12}_{\tau}$ moment, because similarly to $R_{\tau}$ the $D=4$
contribution is suppressed for this moment, and there is significant
contribution from the $D=6$ term. If in the SVZ expansion we retain
only the $D=6$ term, which appears to be a dominant source of the
uncertainty in the nonperturbative sector, then a simplest
self-consistent approach is to take $R_{\tau}$ and
$R^{12}_{\tau}$. This is assumed in the following.

A detailed discussion of the RS dependence of $\delta_{pt}^{tot}$ was
presented  elsewhere~\cite{par-rt}, so in this note we concentrate on
$\delta_{pt}^{12}$. The QCD correction $\delta_{pt}^{12}$ may be
expressed as a contour integral in the complex energy plane with the
so called Adler function under the integral:
\be
\delta_{pt}^{12}=\frac{i}{\pi}\int_{C}\frac{d\sigma}{\sigma}
f^{12}(\frac{\sigma}{m_{\tau}^{2}})\delta_{D,V}(-\sigma),
\label{eq:cont}
\ee
 where $C\,$ is a contour running clockwise from
$\sigma=m^{2}_{\tau}-i{\epsilon}$ to $\sigma=m^{2}_{\tau}+i{\epsilon}$
away from the region of small $|\sigma|$. In the actual calculation we
assume 
that $C$ is a circle $|\sigma|=m_{\tau}^{2}$. The Adler function is
defined by the relation:
\be
(-12\pi^{2}) \sigma \frac{d\,}{d\sigma}
\Pi_{V}^{(1)}(\sigma)=3S_{CKM}[1+\delta_{D}(-\sigma)]
\ee
where $\Pi_{V}^{(1)}$ denotes transverse part of the vector current
correlator. The function $f^{12}(\sigma/m_{\tau}^{2})$ has the form:
\bea
f^{12}(x)&=&\frac{1}{2}-\frac{70}{13}x^{3}+
\frac{105}{26}x^{4} \nonumber \\
         & &+\frac{126}{13}x^{5}-\frac{175}{13}x^{6}+\frac{60}{13}x^{7}.
\eea
The NNLO renormalization group improved perturbative expansion for
$\delta_{D,V}$ may be written in the form:
\begin{equation}
\delta_{D,V}^{(2)}(-\sigma) = a(-\sigma)[1+
r_{1}a(-\sigma)+r_{2}a^{2}(-\sigma)],
\label{del}
\end{equation}
 where $a=g^{2}/(4 \pi^{2})$ denotes the running coupling constant
 that satisfies the NNLO renormalization group equation:
\begin{equation}
\sigma \frac{d\,a}{d\sigma} = - \frac{b}{2}
\,a^{2}\,(1 + c_{1}a + c_{2}a^{2}\,).
\label{rge}
\end{equation}
In the $\overline{MS}$ scheme we have~\cite{r1r2}
$r_{1}^{\overline{MS}}=1.63982$ and $r_{2}^{\overline{MS}}=6.37101$.
The renormalization group coefficients for $n_{f}=3$ are $b=4.5$,
$c_{1}=16/9$ and $c_{2}^{\overline{MS}}=3863/864\approx4.471$.

\begin{figure}
%\center
~\epsfig{file=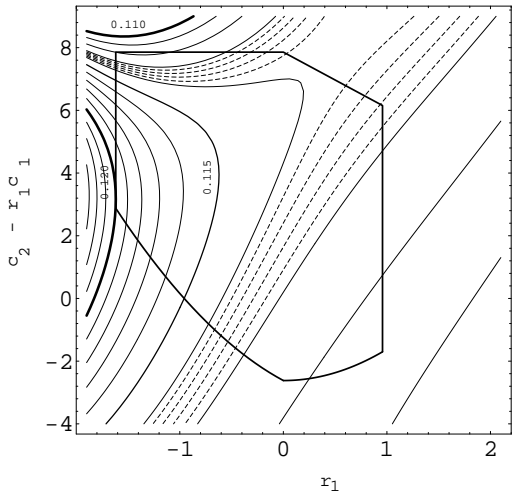,bbllx=209pt,bblly=303pt,bburx=403pt,bbury=488pt} 
\caption{Contour plot of $\delta_{pt}^{12}$ as a function of the
scheme parameters $r_{1}$ and $c_{2}$, for
$\Lambda_{\overline{MS}}^{(3)}=325~\mbox{MeV}$. For technical reasons
we use $c_{2}-c_{1}r_{1}$ on the vertical axis instead of $c_{2}$. The
boundary of the region of scheme parameters satisfying the
Eq.~\protect\ref{allow} is also indicated.}
\label{fig:cplot}
\end{figure}

By keeping the renormalization group improved expression for the Adler
function under the integral and evaluating the contour integral
numerically we obtain an essential improvement of the conventional
perturbation expansion for $\delta_{pt}^{12}$, resumming to all orders
some of the corrections arising from analytic continuation from
spacelike to timelike momenta.~\cite{contour}

The predictions calculated in the next-to-next-to-leading order (NNLO)
approximation depend on two RS parameters, which in principle may be
arbitrary.  Similarly as in the previous work~\cite{par-rt} we parametrize
the freedom of choice of the RS by the parameters $r_{1}$ and $c_{2}$
--- the parameter $r_{2}$ is determined using the RS invariant
combination:
\be
\rho_{2}=c_{2}+r_{2}-c_{1}r_{1}-r_{1}^{2}.
\label{rho2}
\ee
For the Adler function we have $\rho_{2}=5.23783$.  

To optimize the choice of the RS we use the principle of minimal
sensitivity~\cite{pms} (PMS), which singles out the scheme parameters
for which the finite order prediction is least sensitive to the change
of RS, similarly to what we expect from the actual physical quantity.
The dependence of $\delta_{pt}^{12}$ on the scheme parameters is
illustrated in figure~\ref{fig:cplot} for
$\Lambda_{\overline{MS}}^{(3)}=325~\mbox{MeV}$. (For technical reasons
we use $c_{2}-c_{1}r_{1}$ on the vertical axis instead of $c_{2}$.)
We choose as our optimized parameters $r_{1}=0$ and
$c_{2}=1.5\rho_{2}$ --- for small values of
$\Lambda_{\overline{MS}}^{(3)}$ this point lies very close to the
critical point, and even for large values of
$\Lambda_{\overline{MS}}^{(3)}$ the RS dependence in the vicinity of
this point is very small.

To investigate the stability of the predictions we use the condition
proposed by the present author,~\cite{par1} based on the notion that
natural renormalization schemes should not induce extensive
cancellations in the expression for the RS invariant $\rho_{2}$.  The
schemes that have the same --- or smaller --- degree of cancellation
in $\rho_{2}$ as the preferred scheme, satisfy in our case the
inequality:
\be
|c_{2}|+|r_{2}|+c_{1}|r_{1}|+r_{1}^{2}\leq 2\,|\rho_{2}|.
\label{allow}
\ee
The boundary of the relevant region in the scheme parameter space has
been denoted in the figure~\ref{fig:cplot}.  Let us note that the
commonly used $\overline{MS}$ scheme lies outside of this region, but
the numerical value of the prediction is close to the lowest value
obtained within the ``allowed'' region.

\begin{figure}
\center
~\epsfig{file=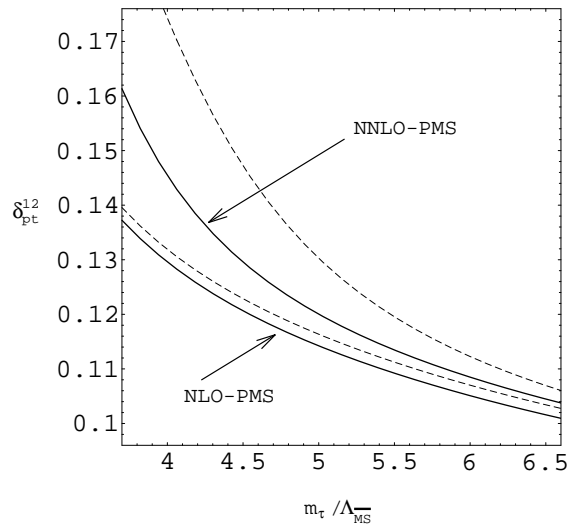,bbllx=202pt,bblly=295pt,bburx=410pt,bbury=497pt} 
\caption{$\delta_{pt}^{12}$ as a function of
$m_{\tau}/\Lambda_{\overline{MS}}^{(3)}$, obtained in the PMS
scheme in NNLO and NLO (upper and lower solid curves, respectively).
The dashed lines indicate variation of the prediction when scheme
parameters are changed within the region satisfying the
Eq.~\protect\ref{allow}.}
\label{fig:endep}
\end{figure}

In figure~\ref{fig:endep} the NNLO PMS predictions for
$\delta_{pt}^{12}$ are shown as a function of
$m_{\tau}/\Lambda_{\overline{MS}}^{(3)}$, together with the minimal
and maximal value obtained by varying the scheme parameters within the
region determined by the condition (\ref{allow}).  We see that the NNLO
predictions for $\delta_{pt}^{12}$, obtained by numerically evaluating
the countour integral expression (\ref{eq:cont}), are free from
potentially dangerous RS instabilities even for large values of
$\Lambda_{\overline{MS}}^{(3)}$. This situation is similar to that
encountered for $\delta_{pt}^{tot}$.  For comparison we also show the
PMS predictions obtained in the next-to-leading order (NLO). (In NLO we
have $r_{1}^{PMS}\approx-0.64$.)  We see that RS dependence of NNLO
expression within the region defined by the condition (\ref{allow}) is
smaller than the difference between NNLO and NLO PMS predictions.

In order to see how the PMS optimization affects the fits to the
experimental data we first test the accuracy of the approximation in
which one only retains the $O_{6}$ contribution in the SVZ
expansion. To this end we make a fit of $\alpha_{s}$ and $O_{6}$ in
the $\overline{MS}$ scheme, and compare the results with the fit
performed by ALEPH~\cite{aleph95}, in which the $O_{4}$, $O_{6}$ and
$O_{8}$ contributions have been taken into account in the (1,0),
(1,1), (1,2) and (1,3) moments. If we take, following ALEPH~\cite{aleph95},
$R_{\tau}=3.645\pm0.024$ and 
$D_{\tau}^{12}=R_{\tau}^{12}/R_{\tau}^{00}=0.0570\pm0.0013$, we obtain
from the fit in the $\overline{MS}$ scheme
$\alpha_{s}(M_{Z}^{2})=0.1209\pm0.0013$ and $O_{6}=-0.0010\pm0.0012$.
This appears to be remarkably close to the values 0.121 and -0.0016
obtained in the full fit by ALEPH. This gives us confidence that the
``$O_{6}$~approximation'' captures the essential features of QCD
corrections in $\tau$ decays.

Performing the same fit, but using now the NNLO PMS predictions, we
obtain $\alpha_{s}(M_{Z}^{2})=0.1198$ and $O_{6}=-0.0011$, i.\ e.\ the
value of the condensate practically does not change, but the value of
the strong coupling constant is reduced by about one standard
deviation of the experimental error. It is also of some interest to
compare these results with the NLO PMS fit --- we then obtain
$\alpha_{s}(M_{Z}^{2})=0.1221$. Taking the difference of the NNLO and
NLO PMS fits is perhaps the best way of estimating the accuracy of the
perturbative prediction. We see that thus obtained uncertainty of
$\alpha_{s}$ is of the order 0.0022, i.\ e.\ it is quite large,
compared for example to the experimental uncertainty.

We may now use the optimized predictions to obtain $\alpha_{s}$
from the more up to date experimental data on $\tau$ decays.  We take
$R_{\tau}=3.613\pm0.032$, which is a weighted average of three
possible determinations involving $B_{e}=0.1790\pm0.0017$,
$B_{\mu}=0.1744\pm0.0023$ and
$\tau_{\tau}=(292.0\pm2.1)\times10^{-15}\mbox{sec}$ (values according
to the
1995 update of the Particle Data Group~\cite{pdg95}). We also take
$D_{\tau}^{12}=0.0561\pm0.0006$, which is a weighted average of the
ALEPH~\cite{aleph95} and CLEO~\cite{cleo} determinations. With these numbers
we obtain $\alpha_{s}(M_{Z}^{2})=0.1177\pm0.0017$
($\alpha_{s}(m_{\tau}^{2})=0.321\pm0.014$) and
$O_{6}=-0.0020\pm0.0005$.

Concluding, the perturbative QCD correction to the $R_{\tau}^{12}$
moment of the invariant mass distribution in hadronic tau decays was
found to be stable with respect to change of the renormalization
scheme, despite the low energy scale, provided that the contour
integral expression is used, which resums some of the corrections to
all orders. However, the difference between predictions in the
conventionally used $\overline{MS}$ scheme and the PMS scheme was
found to be phenomenologically significant.  Also, the difference
between the NNLO and NLO predictions in the PMS scheme was found to be
significant, indicating perhaps that the uncertainty in the
perturbative prediction is larger than previously expected. Finally,
the optimized predictions have been used to obtain a realistic fit for
$\alpha_{s}$ from the experimental data.

%\section*{Acknowledgments}


\section*{References}
\begin{thebibliography}{99}

\bibitem{par1} P.A. R\c{a}czka, {\em Z. Phys.} C {\bf 65},
481 (1995).

\bibitem{aleph93} ALEPH Collab., D. Buskulic {\it et al}, {\em Phys.
Lett.} B {\bf 307}, 209 (1993).

\bibitem{aleph95} P. Reeves, in Proceedings of the XXXth Rencontres de
Moriond Conference ``'95 QCD and High Energy Hadronic Interactions,''
Les Arcs, Savoie, France, 1995, edited by J. Tr\^{a}n Thanh V\^{a}n
(Editions Frontieres, Gif-sur-Yvette, 1995), p.235.

\bibitem{cleo} CLEO Collab., T. Coan {\it et al}, {\em Phys. Lett.} B
{\bf 356}, 580 (1995).

\bibitem{branapi} E. Braaten, {\em Phys. Rev. Lett.} {\bf 60}, 1606
(1988), ibid. {\bf 63}, 577 (1989), S. Narison and A. Pich, {\em
 Phys. Lett.} B {\bf 211}, 183 (1988), E. Braaten, S. Narison and
 A. Pich, {\em Nucl. Phys.}  B {\bf 373}, 581 (1992).

\bibitem{ledepi-kl} F. LeDiberder and A. Pich, {\em Phys. Lett.} B {\bf
289}, 165 (1992).

\bibitem{svz} M.A. Shifman, A.I. Vainshtein and V.I.\\
 Zakharov, {\em Nucl. Phys.} B {\bf 147}, 385, 448, 519 (1979).

\bibitem{par-rt} P.A. R\c{a}czka and A. Szymacha, {\em Z. Phys.} C {\bf 70},
125 (1996).

\bibitem{r1r2} S.G. Gorishny, A.L. Kataev and S.A. Larin,
{\em Phys. Lett.} B {\bf 259}, 144 (1991), M.A. Samuel and
L.R. Surguladze, {\em Phys. Rev.}  D {\bf 44}, 1602 (1991).

\bibitem{contour} A.A. Pivovarov, {\em Z. Phys.} C {\bf 53}, 461 (1992),
 F. LeDiberder and A. Pich, {\em Phys. Lett.} B {\bf 286}, 147 (1992).

\bibitem{pms} P.M. Stevenson,  {\em Phys. Lett.} B {\bf 100}, 61 (1981),
{\em Phys. Rev.} D {\bf 23}, 2916 (1981).

\bibitem{pdg95} Review of Particle Properties, Particle Data Group, L.
Montanet {et al}, 1995 off-year partial update.


\end{thebibliography}
\end{document}